\documentclass[conference]{IEEEtran}

\usepackage{amsmath, amssymb, bm, cite, epsfig, psfrag}
\usepackage{epstopdf}
\usepackage{graphicx}
\usepackage{algorithm,algpseudocode}
\usepackage{array}
\usepackage[margin=10pt,font=small]{caption}
\usepackage{bbm}
\usepackage{multirow}
\usepackage[usenames,dvipsnames]{xcolor}

\usepackage{subfigure}
\usepackage{etoolbox}
\usepackage{pbox}
\usepackage[width=0.48\textwidth]{caption}
\usepackage{colortbl}
\usepackage{fancyhdr}
\newtoggle{conference}
\togglefalse{conference} 
\graphicspath{{figures/}}

\def\beq{\begin{equation}}
\def\eeq{\end{equation}}
\def\beqa{\begin{eqnarray}}
\def\eeqa{\end{eqnarray}}
\def\beqan{\begin{eqnarray*}}
\def\eeqan{\end{eqnarray*}}

\def\PL{\mathrm{PL}}
\def\dB{\mathrm{dB}}

\def\tm1{t\! - \! 1}
\def\tp1{t\! + \! 1}

\def\PL{\mathrm{PL}}
\def\dB{\mathrm{dB}}

\usepackage{lipsum}
\usepackage{fancyhdr}
\usepackage{datetime}

\usepackage{tikz}
\usetikzlibrary{calc}

\begin{document}

\newcommand\blfootnote[1]{%
  \begingroup
  \renewcommand\thefootnote{}\footnote{#1}%
  \addtocounter{footnote}{-1}%
  \endgroup
}
\pagestyle{empty}

\bibliographystyle{IEEEtran}

\title{Probabilistic Omnidirectional Path Loss Models for Millimeter-Wave Outdoor Communications}

\author{
	\IEEEauthorblockN{Mathew K. Samimi, Theodore S. Rappaport and  George R. MacCartney, Jr.}
	mks@nyu.edu, tsr@nyu.edu, gmac@nyu.edu\\
}

\maketitle

\begin{abstract}
This letter presents a probabilistic omnidirectional millimeter-wave path loss model based on real-world 28 GHz and 73 GHz measurements collected in New York City. The probabilistic path loss approach uses a free space line-of-sight propagation model, and for non-line-of-sight conditions uses either a close-in free space reference distance path loss model or a floating-intercept path loss model. The probabilistic model employs a weighting function that specifies the line-of-sight probability for a given transmitter-receiver separation distance. Results show that the probabilistic path loss model offers virtually identical results whether one uses a non-line-of-sight close-in free space reference distance path loss model, with a reference distance of 1 meter, or a floating-intercept path loss model. This letter also shows that site-specific environmental information may be used to yield the probabilistic weighting function for choosing between line-of-sight and non-line-of-sight conditions.
\end{abstract}

 \begin{IEEEkeywords}
 mmWave; close-in free space reference; floating-intercept; probabilistic path loss; ray-tracing; site-specific.
 \end{IEEEkeywords}
\section{Introduction}

\blfootnote{This work was sponsored by National Science Foundation (NSF) CNS CISE Award 1320472, and the NYU WIRELESS Industrial Affiliates.}

Predicting omnidirectional path loss in dense urban millimeter-wave (mmWave) channels is vital for system design and for estimating coverage and capacity of emerging ultrawideband wireless networks~\cite{Pi11}~\cite{Rap15}. Propagation path loss models have been synthesized from the collected unique pointing angle (directional) 28 GHz and 73 GHz mmWave measurements in New York City reported in~\cite{MacCartney14:1}\cite{Rap13:2}, using both the traditional close-in free space reference distance model, and the floating-intercept least-squares regression model~\cite{Rap15}\cite{MacCartney14:2}. 

Here, we present omnidirectional path loss models based on the same New York City data from both line-of-sight (LOS) and non-line-of-sight (NLOS) locations, but consider a site-specific function that describes the probability of having a LOS path for a given transmitter-receiver (T-R) separation distance. In this new ``hybrid'' path loss model, the mean estimated path loss is probabilistic. 

\section{Probability of Line-Of-Sight}

The probability of LOS corresponds to the probability that radiation from the transmitter (TX) will not be blocked by buildings or other obstructions, traveling along a straight and unobstructed propagation path in the urban environment (i.e., zero reflections) to the receiver (RX). Similarly, the NLOS probability corresponds to the probability that the radiation will be obstructed by at least one object, and travel along an obstructed path to reach the RX (i.e., via scattering, or from one or more reflections). These two probabilities heavily depend upon the physical, site-specific environment in which the TX and RX are located.

\begin{figure}
    \centering
 \includegraphics[width=3.5in]{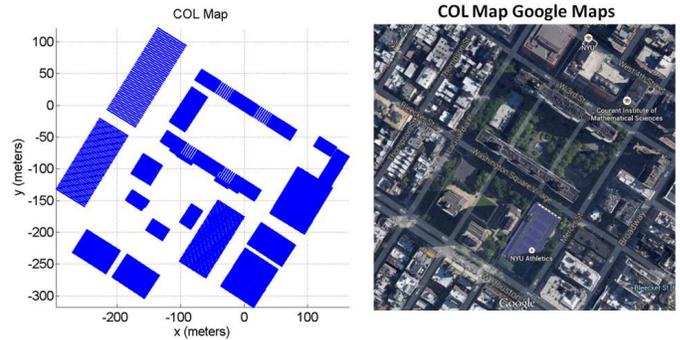}
    \caption{Top view of the Coles (COL) Sports Center environment taken from Google Maps (right), and corresponding environment (top view) reproduced in MATLAB (left). Buildings are represented by blue objects, and the white areas represent free space.}
    \label{fig:map}
\end{figure}

In this work, we obtained the LOS probabilities from ray-tracing techniques. Specifically, all buildings near the transmitters were  represented in a 3-dimensional (3-D) database (Google SketchUp, via Google Maps), which allowed fast and easy 3-D site-specific modeling using simple geometrical shapes such as cubes. The 3-D geometric information was then exported from Google SketchUp into XML format, and subsequently extracted to numerically discretize the environment in MATLAB. For each pair of TX and RX locations, a simple test was performed to determine whether any of the database objects (buildings) blocked the direct connection line (LOS) between the TX and RX. Distances between the TX and RX ranged from 10 m to 200 m, where the base station locations were selected to represent four measured physical TX locations used in the measurements~\cite{MacCartney14:1}\cite{Rap13:2}, and for a RX height of 1.5 m. In the numerical database created to simulate the urban environment of downtown Manhattan, buildings were modeled as 3-D cubes with perfectly smooth, flat surfaces, while smaller obstructions such as trees, lampposts, and vehicular traffic were ignored, as represented in Fig.~\ref{fig:map}.

\begin{figure}
    \centering
 \includegraphics[width=3.5in]{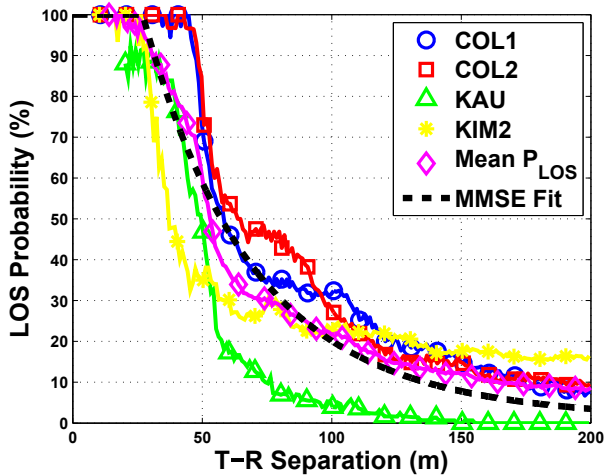}
    \caption{LOS probability curves from ray-tracing as a function of T-R separation distance for TX locations Coles 1 (COL1), Coles 2 (COL2), Kaufman (KAU), and Kimmel 2 (KIM2) obtained from a 3-D site-specific database modeling the downtown Manhattan area in which the measurements were obtained~\cite{MacCartney14:1,Rap13:2}. The mean LOS probability curve was obtained by averaging the four LOS probability curves. The MMSE curve was computed using the mean LOS probability, which yielded $\alpha$ and $d_{\rm BP}$ in~(\ref{Eq1}).}
    \label{fig:LOSProb}
\end{figure}

In order to estimate the LOS probability for a T-R separation distance R, a circle was discretized in 100 evenly-spaced points on the circumference around each TX location in the environment database. For each position along the circle external to a building or obstruction, ray-tracing was used to draw a line from the RX to the TX. If that line to the TX penetrated through at least one building, the corresponding initial position at radius R on the circle was denoted as a NLOS position. If the line to the TX was unobstructed, then that position was counted as a LOS position. This was repeated for all positions along the circle circumference, and the ratio of the number of LOS positions to the number of positions along the circle provided the LOS probability. This was performed over radii ranging from 10 m to 200 m, in increments of 1 m. 

 The LOS probability curves thus obtained for the New York City measurements collected in~\cite{MacCartney14:1,Rap13:2} at transmitter sites Coles 1 (COL1), Coles 2 (COL2), Kaufman (KAU), and Kimmel 2 (KIM2) are shown in Fig.~\ref{fig:LOSProb}. Previous work considered the probability of LOS based on actual measured RX locations~\cite{Akdeniz14}, while in this work, the LOS probability is determined from ray-tracing simulations that consider the universe of all possible locations at the New York City locations. Note that the fifth measured TX location at Kimmel 1 (KIM1) was disregarded for modeling the LOS probability, because it was located near Washington Square Park, a large open area park, resulting in unusually large LOS probabilities as far as 150 m away from the base station. As the T-R separation distance increases from 10 m to about 30 m, the probability of LOS remains constant with a value of 100\%, and decreases monotonically after 30 m, as the environment becomes denser with more path obstructions. The mean LOS probability was computed from the four distinct LOS probability curves from the four physical TX locations used in~\cite{MacCartney14:1} and~\cite{Rap13:2}. The mean LOS probability curve was fit to an analytical function of the form:
\begin{equation}\label{Eq1}
P_{\rm LOS}(d) = \left[\min \left(\frac{d_{\rm BP}}{d},1 \right)\left( 1- e^{-\frac{d}{\alpha}}  \right)+e^{-\frac{d}{\alpha}}\right]^2
\end{equation}

\noindent where $d_{\rm BP}$ is the breakpoint distance at which the LOS probability is no longer equal to 1, and $\alpha$ (m) is a decay parameter. In this work, we applied the minimum mean square error (MMSE) method, which yielded values of $d_{\rm BP} = 27$ m and $\alpha=71$ m that minimize the mean square error between the mean LOS curve in Fig.~\ref{fig:LOSProb} and (\ref{Eq1}). Table~\ref{tbl:singleTX} shows the different $d_{BP}$ and $\alpha$ values for each individual TX considered in this work. Other cities and environments will likely have different values of $d_{\rm BP}$ and $\alpha$, based on the density of buildings, the width of streets, and the heights of TX and RX antennas. Note that the WINNER probability function in the LOS microcellular environment~\cite{WinnerII} uses the same form as in~(\ref{Eq1}) but without the square exponent, which yielded a greater error to the mean LOS curve (in Fig.~\ref{fig:LOSProb}) than~(\ref{Eq1}).

\begin{table}
\centering
\caption{Summary of $d_{BP}$ and $\alpha$ values parameterizing (\ref{Eq1}), obtained from the MMSE method, for Coles 1 (COL1), Coles 2 (COL2), Kaufman (KAU), and Kimmel 2 (KIM2) TX locations considered in this work.}

\resizebox{9cm}{!} {

\begin{tabular}{|c|c|c|c|c|c|}

\hline
\textbf{TX ID} 	& \textbf{Latitude}  	& \textbf{Longitude} 	& \textbf{Height (m)}	& $\bm{d_{BP}}$ \textbf{(m)}	& $\bm{\alpha}$ \textbf{(m)} \tabularnewline \hline

COL1	&	40.7270944	&	-73.9974972 	& 7	& 36	& 71 \tabularnewline \hline

COL2 & 	40.7268833	&	-73.9970556		& 7	& 39	& 68 \tabularnewline \hline

KAU	&	40.7290611	&	-73.9962500		& 17	& 30	& 21 \tabularnewline \hline

KIM2	& 	40.7297444	& 	-73.9977222		& 7	& 15	& 95 \tabularnewline \hline

All	&	- 		& - 				& 7 - 17	& 27	& 71 \tabularnewline \hline

\end{tabular}

}
\label{tbl:singleTX}
\end{table}

\section{Probabilistic Path Loss Model}
\label{sec:PathLoss}

The hybrid probabilistic path loss model offers an alternative to conventional propagation path loss modeling over distance. Currently, two popular propagation modeling approaches include fitting a measured path loss data set over a model that uses the close-in free space reference distance and floating-intercept path loss models~\cite{Rap15}. The floating-intercept model provides a method for estimating path loss data in a given range of measured T-R separations, but can give non-realistic path loss results if extrapolated outside the measured range. It is important to note that the slope of the floating-intercept model often has no physical basis. The close-in free space reference distance model is physically based and adequately estimates path loss data points, but is sensitive to the selected free space reference distance anchor point $d_0$ when estimating the NLOS data, where the choice of $d_0$ is subjective. Establishing a standard free space reference distance of  $d_0=1$  m for all mmWave measurements and path loss models removes this subjectivity, and offers a standard approach for propagation models at any mmWave frequency with any antenna. As long as measurements are obtained in the far field of an antenna, the measured data and corresponding path loss models may be recast with a 1 m reference distance. This is particularly valuable when comparing propagation measurements over different mmWave frequencies, since the biggest difference in propagation path loss at mmWave frequencies has been shown to be in the first meter of propagation~\cite{Rap15}.

MmWave frequencies, given their small wavelengths, are much more sensitive to whether LOS or NLOS conditions exist, and future systems are likely to use highly directional antennas that search for energy over all directions. By bringing both LOS and NLOS models together to improve path loss estimation, we propose here to combine LOS and NLOS propagation using the close-in free space reference distance model (for LOS) and a floating-intercept model (for NLOS), respectively, and weigh the LOS and NLOS path losses using a probabilistic distribution for the probability of LOS as a function of T-R separation. We also estimate the probabilistic path loss by combining the LOS and NLOS close-in reference distance path loss models with respect to a fixed reference distance of $d_0=1$ m, and the probability of LOS shown in (\ref{Eq1}). Establishing a fixed reference distance of $d_0=1$ m is also a viable approach for both LOS and NLOS modeling, and we validate our new probabilistic propagation modeling scheme to demonstrate the flexibility of our approach, while showing the inherent simplicity (with very little loss of accuracy) of using a simple 1 m close-in free space reference distance model for both LOS and NLOS environments. 

The LOS and NLOS path loss equation lines used in this letter, and previously published in~\cite{Rap15,MacCartney14:2}, are of the form:
\begin{equation}\label{GenForm}\small\begin{split}
PL[\dB](d) = 20\log_{10}\left(\frac{4\pi d_0}{\lambda}\right)&+10\overline{n}\log_{10}\left(\frac{d}{d_0}\right)+X_{\sigma}\\
& \hspace{3cm}d\geq d_0
\end{split}\end{equation}
\noindent where $d_0=1$ m, and it follows that:
\begin{equation}\label{LOSeq}\small\begin{split}
PL_{\mathrm{LOS}}[\dB](d) = 20\log_{10}\left(\frac{4\pi}{\lambda}\right)&+10\overline{n}_{\mathrm{LOS}}\log_{10}(d)+X_{\sigma,\mathrm{LOS}}\\
&   \hspace{2.8cm}d\geq 1\text{ m}\\
\end{split}\end{equation}
\begin{equation}\label{NLOSCloseInEq}\begin{split}
PL_{\mathrm{NLOS,Close-In}}&[\dB](d) = 20\log_{10}\left(\frac{4\pi}{\lambda}\right)+\\
&10\overline{n}_{\mathrm{NLOS}}\log_{10}(d) + X_{\sigma,\mathrm{NLOS}}, \hspace{.3cm}d\geq 1\text{ m}
\end{split}\end{equation}
\begin{equation}\label{NLOSFloatingEq}\begin{split}
PL_{\mathrm{NLOS,Floating}}[\dB](d) = \alpha +10\beta&\log_{10}(d)+X_{\sigma,\mathrm{NLOS}}\\
&\hspace{.6cm}30 \text{ m} < d < 200 \text{ m}
\end{split}\end{equation}

\noindent where $PL_{\mathrm{LOS}}$ is the LOS free space path loss, $PL_{\mathrm{NLOS,Close-In}}$ and $PL_{\mathrm{NLOS,Floating}}$ are the NLOS path losses computed using the 1 m close-in free space reference distance and the floating-intercept models, respectively, $\lambda$ is the carrier wavelength, $\overline{n}_{\mathrm{LOS}}$ and $\overline{n}_{\mathrm{NLOS}}$ are the average (over distance) path loss exponents in LOS and NLOS, respectively, $\alpha$ and $\beta$ are the intercept and slope of the floating-intercept model parameters, and $X_{\sigma}$ is the lognormal random variable (normal in dB) with standard deviation $\sigma$ (dB) to model large-scale shadowing. 

As found in~\cite{Rap15,MacCartney14:2}, the omnidirectional LOS path loss exponent and shadowing factor with respect to a 1 m free space reference distance were computed to be $\overline{n}_{\mathrm{LOS}}=2.1$ and $\sigma_{\mathrm{LOS}} = 3.6$ dB at 28 GHz, and $\overline{n}_{\mathrm{LOS}}=2.0$ and $\sigma_{\mathrm{LOS}}=4.8$ dB at 73 GHz, respectively, indicating a close match with true free space propagation ($n=2$). The omnidirectional NLOS path loss exponent and shadowing factor with respect to a 1 m free space reference distance were found in~\cite{Rap15,MacCartney14:2} to be $\overline{n}_{\mathrm{NLOS}}=3.4$ and $\sigma_{\mathrm{NLOS}} = 9.7$ dB at 28 GHz, and $\overline{n}_{\mathrm{NLOS}}=3.4$ and $\sigma_{\mathrm{NLOS}}=7.9$ dB at 73 GHz, respectively, indicating greater signal attenuation over distance as compared to LOS conditions. The NLOS floating-intercept path loss equation lines produced the following parameters: $\alpha=79.2$ dB, $\beta=2.6$ and $\sigma_{\mathrm{NLOS}}=9.6$ dB at 28 GHz, and  $\alpha=80.6$ dB, $\beta=2.9$ and $\sigma_{\mathrm{NLOS}}=7.8$ dB at 73 GHz~\cite{Rap15,MacCartney14:2}. 

\begin{figure}
    \centering
 \includegraphics[width=3.5in]{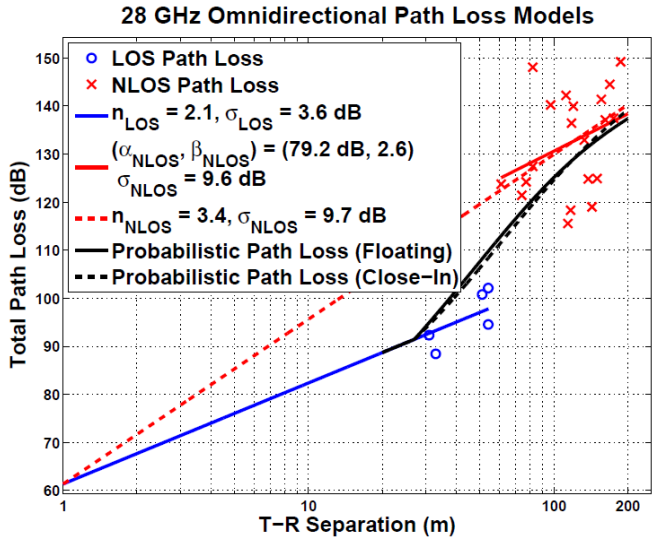}
    \caption{28 GHz omnidirectional path loss models obtained from wideband propagation measurements in New York City~\cite{Rap13:2}. The LOS situation is modeled using the close-in free space reference distance model with respect to 1 m, and the NLOS situation is modeled using both the 1 m close-in free space reference distance and the floating-intercept models, showing virtually no difference in the resulting probabilistic path loss equation lines, obtained from (\ref{Eq:3}).}
    \label{fig:28GHzOmni}
\end{figure}

\begin{figure}
    \centering
 \includegraphics[width=3.5in]{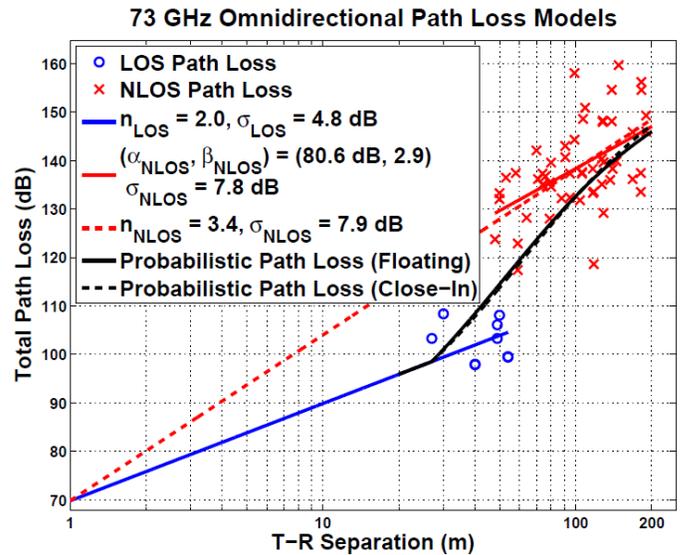}
    \caption{73 GHz omnidirectional path loss models obtained from wideband propagation measurements in New York City, where different RX antenna heights are used as described in~\cite{MacCartney14:1}, according to Eqs. (\ref{LOSeq})-(\ref{Eq:3}). Note that the probabilistic path loss model yields virtually identical results using either a 1 m close-in free space reference distance or a floating-intercept path loss model for NLOS conditions.}
    \label{fig:73GHzOmni}
\end{figure}

Fig.~\ref{fig:28GHzOmni} and Fig.~\ref{fig:73GHzOmni} show the 28 GHz and 73 GHz omnidirectional path loss scatter plots and corresponding mean path loss equation lines, where the LOS path loss line was obtained using the close-in free space reference distance path loss model with respect to a 1 m free space reference distance, and the NLOS path loss lines were obtained using the 1 m close-in free space reference distance model and floating-intercept path loss model. As presented in~\cite{Rap13:2,MacCartney14:2}, 74 TX-RX location combinations were measured at both 28 GHz and 73 GHz, of which 13 and 5 locations provided no measurable path loss, respectively, over T-R distances of 200 m. Using Eqs.~(\ref{LOSeq}),  (\ref{NLOSCloseInEq}), and~(\ref{NLOSFloatingEq}), in conjunction with (\ref{Eq1}), it is possible to implement a general probabilistic path loss model, where in this letter we show, by example, one approach that uses the 1 m close-in free space reference distance in LOS conditions and the NLOS floating-intercept path loss model, whereas the other approach uses the 1 m close-in free space reference distance models in both LOS and NLOS conditions, as shown in (\ref{Eq:3}) and (\ref{Eq:4}):
\begin{equation}
\begin{split}\label{Eq:3}
\PL_{\mathrm{Prob}}[\dB](d) = P_{\mathrm{LOS}}(d)\times \PL_{\mathrm{LOS}}(d)\\ + (1-P_{\mathrm{LOS}}(d))\times \PL_{\mathrm{NLOS}}(d)
\end{split}
\end{equation}

\noindent where $P_{\mathrm{LOS}}(d)$, $\PL_{\mathrm{LOS}}(d)$, and $\PL_{\mathrm{NLOS}}(d)$ are given in Eqs.~(\ref{Eq1}),~(\ref{LOSeq}), and~(\ref{NLOSCloseInEq}) or (\ref{NLOSFloatingEq}), respectively. It is clear that any other distance-dependent path loss model, such as the Stanford University Interim (SUI) model~\cite{Sulyman14}, or other propagation models, could be used here. Eq.~(\ref{Eq:3}) can be rewritten as:
\begin{equation}
\begin{split}\label{Eq:4}
\PL_{\mathrm{Prob}}[\dB](d) = \overline{PL}_{\rm Prob}(d)+ X_{\sigma}(d)
\end{split}
\end{equation}
\noindent where,
\begin{align}
&\overline{PL}_{\rm Prob}(d) = P_{\mathrm{LOS}}(d) \overline{PL}_{\rm LOS}(d)+P_{\mathrm{NLOS}}(d) \overline{PL}_{\rm NLOS}(d)\\ \label{Eq:Shadow}
& X_{\sigma}(d) =  P_{\mathrm{LOS}}(d)  X_{\sigma,\rm LOS}+ P_{\mathrm{NLOS}}(d)  X_{\sigma,\rm NLOS}
\end{align}

\noindent where  $\overline{\PL}_{\mathrm{LOS}}(d)$, and $\overline{\PL}_{\mathrm{NLOS}}(d)$ are the mean LOS and NLOS distance-dependent path loss equations from Eqs.~(\ref{LOSeq}), and~(\ref{NLOSCloseInEq}) or~(\ref{NLOSFloatingEq}), and $X_{\sigma}(d)$ is the sum of two independent 0 dB mean lognormal random variables, also with 0 dB mean, and a distance-dependent standard deviation, i.e., shadow factor, (in dB) \text{$\sigma(d) = \sqrt{P_{\mathrm{LOS}}^2(d)  \sigma^2_{LOS}+ (1-P_{\mathrm{LOS}}(d))^2  \sigma_{NLOS}^2}$}. 

 The probabilistic omnidirectional path loss model shown in (\ref{Eq:3}), and plotted in Figs.~\ref{fig:28GHzOmni} and~\ref{fig:73GHzOmni}, combines both the combination of the close-in reference LOS free space path loss and the floating-intercept NLOS path loss models, and the close-in reference LOS and NLOS path loss model. Fig~\ref{fig:28GHzOmni} shows virtually no difference between the two probabilistic curves, indicating little difference between using the NLOS 1 m close-in free space reference or floating-intercept path loss models. In Figs.~\ref{fig:28GHzOmni} and \ref{fig:73GHzOmni}, the mean probabilistic path loss equation line is plotted, but the standard deviations $\sigma_{\mathrm{LOS}}$ and $\sigma_{\mathrm{NLOS}}$ from the lognormal distributions as shown in Eq.~(\ref{Eq:Shadow}) must also be taken into account when performing system-wide simulations to model large-scale shadowing. The mean of the probabilistic path loss equation (Eq.~(\ref{Eq:3})) is always found between the LOS and NLOS models. Thus, as the T-R separation increases from 10 m to 27 m, the LOS probability remains 100\%, and the probabilistic path loss line follows the LOS measured path loss data. Similarly, as the T-R separation increases from 27 m to 200 m, the LOS probability falls from 100\% to 4\%, respectively.

\section{Conclusion}

This letter presented probabilistic path loss models based on LOS and NLOS omnidirectional propagation path loss data, using a probability distribution for LOS obtained from a 3-D site-specific database in New York City. The probabilistic model was shown to be very similar when using either the NLOS 1 m close-in free space reference distance path loss model, or the floating-intercept path loss model, indicating little to no difference between the two. The probabilistic path loss models given here may be used to estimate signal coverage, interference, and outage as a function of distance, and provide a convenient way to model path loss for future mmWave systems, where highly directional antennas and the smaller wavelengths will be more sensitive to whether LOS conditions exist or not.

\section{Acknowledgment}

The authors wish to thank Professor Andreas F. Molisch and Dr. Carl Gustafson for valuable discussions, and the NYU WIRELESS Industrial Affiliate sponsors, with special thanks to Samsung and Intel Corporation.

\bibliographystyle{IEEEtran}
\bibliography{Samimi_WCL2015-0120.R1}
\end{document}